\documentclass[pra,aps,epsf,twocolumn,notitlepage,showpacs,10pt]{revtex4-1}
\usepackage{graphicx,amsfonts,times,bm,amsmath,amssymb,verbatim,color,array,dsfont,multirow,mathrsfs}

\newcommand{\bra}[1]{\langle {#1} |}
\newcommand{\ket}[1]{| {#1} \rangle}

\newcommand{\<}{\langle}

\renewcommand{\>}{\rangle}
\renewcommand{\(}{\left(}
\renewcommand{\)}{\right)}
\renewcommand{\[}{\left[}
\renewcommand{\]}{\right]}

\begin{document}

\title{Direct Characterization of Quantum Dynamics with Noisy Ancilla}
\author{Eugene Dumitrescu$^{1,2}$, Travis S. Humble$^{1,2}$}

\affiliation{ $^1$Quantum Computing Institute, Oak Ridge National Laboratory, Oak Ridge, TN 37831\\
$^2$ Bredesen Center for Interdisciplinary Research, University of Tennessee, Knoxville, TN 37996}

\begin{abstract}
We present methods for the direct characterization of quantum dynamics (DCQD) in which both the principal and ancilla systems undergo noisy processes. 
Using a concatenated error detection code, we discriminate between located and unlocated errors on the principal system in what amounts to filtering of ancilla noise. 
The example of composite noise involving amplitude damping and depolarizing channels is used to demonstrate the method, while we find the rate of noise filtering is more generally dependent on code distance. Our results indicate the accuracy of quantum process characterization can be greatly improved while remaining within reach of current experimental capabilities. 
\end{abstract}

\maketitle

\section{Introduction}
Decoherence and noisy dynamics in open quantum systems are major hurdles to the realization of working quantum computers and practical quantum communication devices \cite{Nielsen}. 
Information encoded in a controlled quantum system will leak into its surrounding environment when there is coupling between the two systems, resulting in shorter qubit lifetimes and lower gate fidelities. 
It is necessary to characterize these quantum dynamics in order to better understand the sources of system-environment coupling. 
Characterization of the dynamical process can then be used to mitigate sources of noise and improve qubit coherence times. 
\par
Given the density matrix $\rho$ for a $d$-dimensional system, the completely-positive quantum process
\begin{equation}
\label{eq:PM}  
\mathcal{E}(\rho) = \sum_{a} {E_a \rho E_a^{\dagger}} = \sum_{m,n=0}^{d^2-1}{\chi_{mn} F_m \rho F_n^{\dagger}}
\end{equation}
may be described in terms of its Kraus operators $\{E_{a}\}$ or the Hermitian process matrix $\chi$ defined with respect to the operator basis $\{F_m\}$. 
Experimental characterization of the process matrix $\chi$ provides a concrete representation of $\mathcal{E}$ that can be used to study and refine system behavior.
In standard quantum process tomography (SQPT), measurements characterizing the state $\mathcal{E}(\rho)$ are used to reconstruct the process matrix by inverting Eq.~(\ref{eq:PM}) over a complete set of input states \cite{Poyatos, Chuang_97, Nielsen, O_Brien, Bialczak, Kim}. 
Ancilla-assisted process tomography (AAPT) performs a similar inversion using fewer input states by exploiting correlations between the principal system ({\bf P}) and an ancilla system ({\bf A}) isolated from non-trivial quantum process \cite{Altepeter}. 
Such a composite process can be written as $\mathcal{E}_{\bf P} \otimes \mathds{1}_{\bf A} (\rho)$ where the subscripts indicate which processes occur on each subsystem.
\par
In contrast, direct-characterization of quantum dynamics (DCQD) avoids inverting Eq.~(\ref{eq:PM}) by measuring the process elements $\chi_{mn}$ directly \cite{Mohseni_PRL,Mohseni_PRA}. 
DCQD techniques have recently been applied to characterize trapped ion \cite{Nigg_PRL} and hyper-entangled photon \cite{Graham_PRL} dynamics.
Like AAPT, the principal and ancilla subsystems are initially entangled in a probe state before being subjected to non-trivial and trivial quantum processes respectively. 
Interestingly, DCQD probe states can be described as the codewords of a quantum error correction (QEC) code. 
In this framework, a quantum process maps the joint system probe state either within or outside the codespace. 
Processes mapping the probe state outside the original codespace are detected and characterized by their error syndrome, i.e., the measured eigenvalues of each QEC code generators. 
Syndrome frequencies derived from an ensemble of stabilizer measurements are sufficient to directly characterize the underlying process matrix $\chi$.
\par
The DCQD framework shows that the mathematical tools developed for QEC can be leveraged for process characterization. 
In particular, code design plays an important role in probing the process matrix \cite{Mohseni_PRL,Mohseni_PRA}. 
Recent extensions to DCQD involve generalized characterization codes which also encode logical quantum information \cite{Omkar1,Omkar2}.
This offers the ability to characterize processes occurring during an arbitrary quantum computation.
Complimentary works, from a QEC perspective, have shown that syndrome data generated by error correction protocols can be used for noisy parameter estimation \cite{Combes2014,Fujiwara2014,Fowler}, with
recent experiments involving stabilizer QEC circuits over 9 and 4 qubits are prime candidates for these types of characterization methods \cite{Martinis_15,IBM_15}. 
\par
Despite advances within the DCQD paradigm, a significant and persistent drawback in all existing schemes is the requirement that the ancilla system be perfectly noiseless. 
This assumption is necessary for correctly interpreting the measured syndromes in the context of Eq.~(\ref{eq:PM}).
Noisy ancilla lead to spurious data that corrupts the process tomography and adds errors to the process matrix.
However, noise is certainly present in any realistic experiment and it is important to ask how QEC-based process characterization can be extended to include noisy ancilla.
\par
We address the use of noisy ancilla for process characterization by introducing a new class of quantum process codes that remove the requirement of noise-free ancilla.
Our approach is based on concatenated encoding of the ancilla system using a second quantum error detection code.
We show that by monitoring syndrome values of the composite code measurements of the principal system that have been corrupted by ancilla noise can be filtered out. 
By removing measurements attributed to noisy ancilla, we generate a higher fidelity construction of the process matrix than possible with direct characterization alone.
We also examine the question of efficiency, which we define as a tradeoff between the syndromes collected and the accuracy of the process characterization.
\par
The remainder of the paper is organized as follows:
in Sec.~\ref{sec:CDCQD} we review notation for stabilizer QEC codes and outline the conventional DCQD procedure for constructing the process matrix before introducing a concatenated six-qubit code used to characterized the dynamics of a two-qubit principal system in the presence of full system noise.
This includes a discussion of how ancilla error detection is used to filter tomographic information prior to characterization.  
In Sec.~\ref{sec:Monte-Carlo} we present a numerical case study of an amplitude damping channel on various codes with and without depolarizing noise affecting the ancilla subsystem. 
We discuss results of our simulation, the degree to which the code faithfully characterizes dynamics on the principal system, and the probability that high weight errors, which can pass through our concatenated error filter thus corrupting the tomographic data, occur in Sec.~\ref{sec:analysis}. 
Our conclusions and discussion appear in Sec.~\ref{sec:conclusion}.
\section{Characterization with Noisy Ancilla}
\label{sec:CDCQD}
An $[[n,k,d]]$ stabilizer code maps $k$ logical qubits onto $n$ qubits with a distance $d$ between distinct codewords \cite{Gottesman_97}. 
Let $\mathcal{S} = \< g_1, \cdots, g_r \>$ denote an Abelian stabilizer group whose $r=n-k$ generators are drawn from the $n$-qubit Pauli group, i.e., $g_i \in \mathcal{P}_n$.
These stabilizer generators define a set of commuting observables called the syndrome that partitions the $n$-qubit Hilbert space into a set of mutually orthogonal subspaces, each encoding $k$ qubits. 
The $i$-th subspace $\mathscr{H}_i$ corresponds to a syndrome eigenvalue $e_i$.
We will represent each syndrome as a string of classical bits such that $e_{ij} = 0$ or $1$, respectively, for the $+1$ or $-1$ eigenstate of the generator $g_j$.
In this notation, the logical codespace $\mathscr{H}_0$ corresponds to the trivial syndrome $e_0$ generated by the stabilizer group $\mathcal{S} = \{s \in \mathcal{P}_n  :  s \ket{\psi} = \ket{\psi}, \; \forall \ket{\psi} \in \mathscr{H}_0 \}$. 
\par
Errors due to a set of operators $\mathds{E}$ are said to be correctable if the QEC condition
\begin{equation}
\label{eq:QECC}
\bra{i} E_a^\dagger E_b \ket{j} =C_{ab} \delta_{ij}
\end{equation}
is satisfied for all $E_{a},E_b \in \mathds{E}$, where $\ket{i},\ket{j}$ are orthonormal basis vectors spanning $\mathscr{H}_0$ and $C_{ab}$ is a Hermitian matrix.
In particular, a correctable error $E_a$ maps states in the codespace $\mathscr{H}_0$ to another codespace.
Upon measuring a syndrome value $e_i$, a state $\ket{\psi}$ is projected into the subspace $\mathscr{H}_i$. 
For purposes of error correction, the syndrome dictates what recovery operation should be applied to return the state to the logical codespace $\mathscr{H}_0$.
\par
In the context of quantum process characterization the operator set $\mathds{E}$ represents a basis for the dynamical processes by which the encoded state evolves. 
Instead of correcting $\mathds{E}$, the goal of process characterization is to unambiguously detect these operations. 
As such, there is a significant difference between how QEC and process characterization are affected by undetectable errors.
In particular, operators commuting with the stabilizer group belong to the normalizer group $\mathcal{N}(\mathcal{S})$, where elements of the normalizer act as logical operators that map one codestate to another or stabilize the state (since $\mathcal{S} \in \mathcal{N}(\mathcal{S})$).
Elements of $\mathcal{N}(\mathcal{S})$ yield trivial syndromes and lead to logical errors because no recovery operation is applied.
For process characterization, however, undetectable stabilizer and normalizer operators lead to faulty tomographic data as they cannot be distinguished from the identity operation. 
As seen below, concatenated ancilla codes ensure {\em all} weight one errors, across the entire system, are detectable via error syndromes distinct from $e_i$. 
\par
\begin{figure}[tbp]
\begin{center}
\includegraphics[width=\columnwidth]{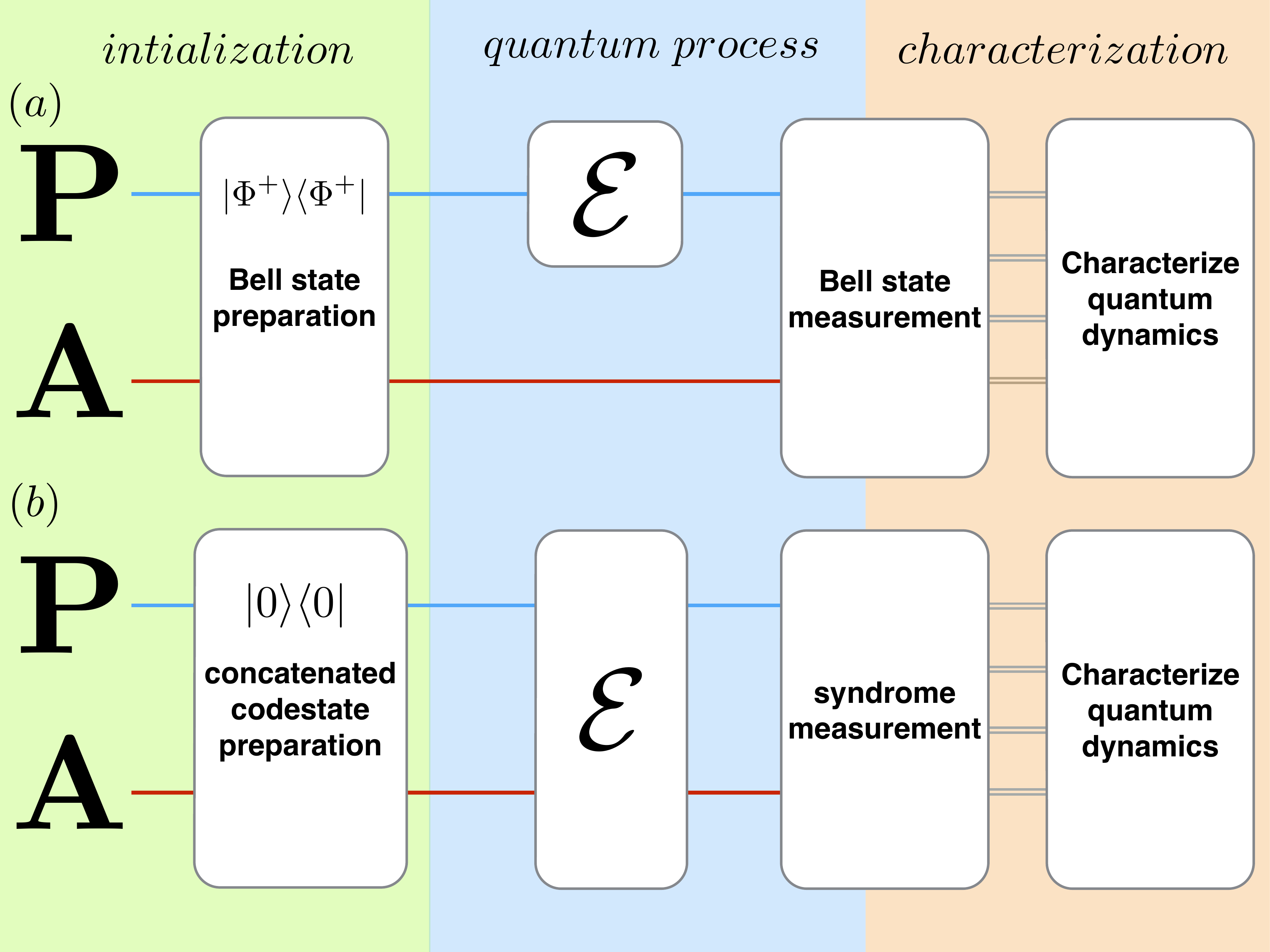}
\caption{\label{fig:schematic}
Outline of a quantum circuit used to directly characterize the dynamics of a principal system {\bf P}. 
The principal system is first entangled with the ancilla system {\bf A} before being subject to some noisy dynamics. 
The process matrix is then constructed from an ensemble of stabilizer measurements (for details see Sec.~\ref{sec:cleandcqd} and Fig.~\ref{fig:DCQD_flowchart}).
Panel (a) illustrates a conventional DCQD circuit, where the {\bf P} and {\bf A} are maximally entangled in a Bell state $\ket{\Phi^+}$ and characterization assumes a noiseless ancilla.
Panel (b) illustrates a concatenated DCQD circuit, with an initial entangled state $\ket{0}$ (Eq.~\ref{eq:codeket}), which supports errors on the ancilla subsystem. }
\end{center}
\end{figure}

DCQD involves partitioning the $n$-qubit system into a $n_{\bf P}$-qubit principal subsystem {\bf P} and a $n_{\bf A}$-qubit ancilla subsystem {\bf A}. 
After the two subsystems are initially entangled, the principal system is subjected to dynamics $\mathcal{E}$ while the ancilla system has previously been assumed to remain isolated from any such process. 
Syndrome measurements project the state into one of the $2^r$ subspaces $\mathscr{H}_i$ defined by the QEC code, and the relative frequency with which each syndrome is observed characterizes the process matrix $\chi$.
A schematic circuit expressing this partition for DCQD is shown in Fig.~\ref{fig:schematic}(a). 

\subsection{Clean Ancilla DCQD}
\label{sec:cleandcqd}
\begin{figure}[t]
\begin{center}
\includegraphics[width=\columnwidth]{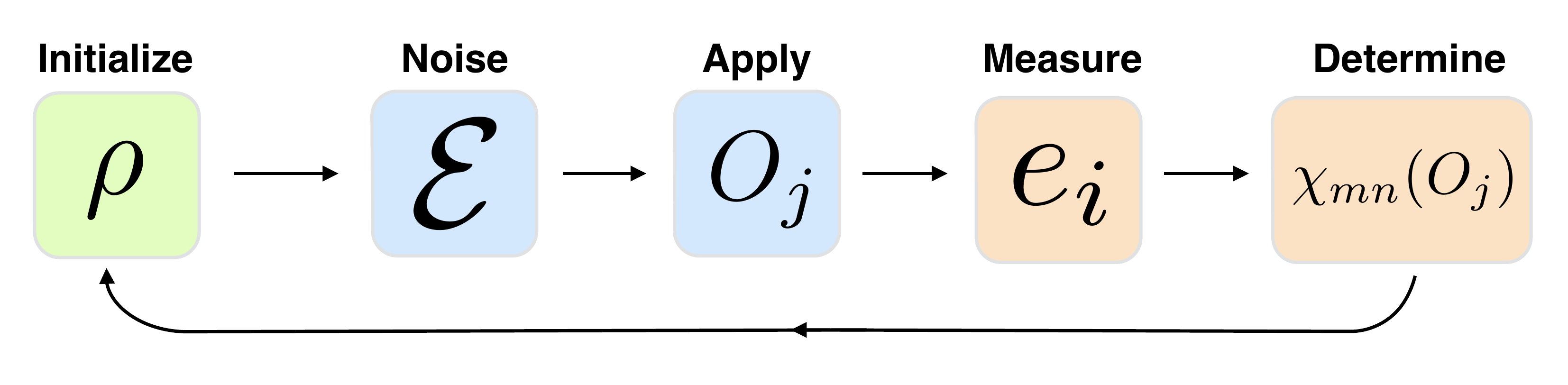}
\caption{\label{fig:DCQD_flowchart}
Schematic of the DCQD process.
Begin with the state $\rho = \ket{0}\bra{0}$ initialized in the codespace of $\mathcal{S}_1$.
Next, subject $\rho$ to some dynamics resulting in $\mathcal{E}(\rho)$. 
A pre-processing operation $O_j= \{\mathds{I}, U_j, P_j\}$ is then applied just prior to the stabilizer generators yielding an error syndrome $e_i$. 
An ensemble of syndrome measurements given the pre-processing $O_j$ is used to deduce part of the process matrix $\chi_{mn}(O_j)= \{\chi_{ii}, \text{Im} \chi_{Ji}, \text{Re} \chi_{Ji}\}$ as described in Eqs.~\ref{eq:syn_prob_diag},\ref{eq:syn_prob_coherence},\ref{eq:syn_prob_coherence_2}.}
\end{center}
\end{figure}
The probability for a state $\rho \equiv \ket{\psi}\bra{\psi}$ to be projectively measured into the subspace $\mathscr{H}_i$ with the error syndrome $e_i$ is 
\begin{equation}
\label{eq:pi}
p_i=\text{Tr} \[ \Pi_i \rho \]
\end{equation}
where $\Pi_i$ is the projector onto $\mathscr{H}_i$. 
We assume a one-dimensional codespace and write the projector into each orthogonal subspace in the {\em stabilizer basis} as $\Pi_i=\ket{i}\bra{i}$ where $\ket{i} = E_i \ket{0}$ for the unique operator $E_i \in \mathds{E}$. 
(Our results will also hold when the syndrome subspaces are $j=2^k$ dimensional and the subspace projectors are $\Pi_i=\sum_j \ket{i_j}\bra{i_j}$.)
Let the encoded state of the system be initialized as $\rho_L=\ket{0}\bra{0}$, so that
before any quantum operation is applied the trivial syndrome $e_0$ occurs with probability $p_0=1$. 
\par
After $\mathcal{E}$ acts on {\bf P}, the probability for each error syndrome becomes \cite{Mohseni_PRL,Omkar1}
\begin{eqnarray} \label{eq:syn_prob_diag}
p_i&=&\text{Tr}\[ \Pi_{i} \sum_{mn} \chi_{mn} F_m \rho_L F_n^\dagger \] 	\\ \nonumber
& = &\chi_{ii}
\end{eqnarray}
where we use the QEC condition in Eq.~(\ref{eq:QECC}) and perform the trace in the stabilizer basis. 
Therefore, the quantum dynamical populations (diagonal elements) of $\chi$ are simply the relative frequencies with which each syndrome $e_i$ appears in an ensemble of stabilizer generators measurements. 
\par
Off-diagonals of $\chi$ represent quantum dynamical coherence and are similarly measured by first applying the unitary operation $U_{j} = (\mathds{1} + i F_j) / \sqrt{2}$, where $F_j \in \mathds{E} $ is a member of the Pauli group with a trivial Pauli phase factor of $+1$. The corresponding probability for each syndrome measurement is then 
\begin{eqnarray} 
\label{eq:syn_prob_coherence}
p_i(U_j)&=& \text{Tr} \[ \Pi_i U_j \mathcal{E} (\rho) U_j^\dagger\] \\ \nonumber
& = & \frac{\chi_{ii} +\chi_{JJ}}{2} -\text{Im} \(\phi_J \chi_{Ji} \)
\end{eqnarray}
where $\phi_{J} \in \{\pm 1, \pm i\} $ is a Pauli phase factor which, along with $F_J$, depends on the indices $i,j$ according to $\phi_J F_J = F_i^\dagger F_j $ \cite{Omkar1}.
\par 
Applying the unitary $U_j$ enables measurements to probe either the real or imaginary part of $\chi_{ij}$.
The complimentary part of a given element $\chi_{ij}$ is recovered by applying instead the projective measurement $P_{j \pm}=\mathds{1} \pm F_j$.
This corresponds to measuring an eigenvalue of $\pm1$ for the operator $F_j$. 
In this case the syndrome probabilities become
\begin{eqnarray} \label{eq:syn_prob_coherence_2}
p_i(P_{j \pm})&=& \text{Tr} \[ \Pi_i P_{j \pm} \mathcal{E} (\rho) P_{j \pm}^\dagger\] \\ \nonumber
& = & \frac{\chi_{ii} +\chi_{JJ}}{2}  \pm \text{Re}\( \phi_J \chi_{Ji}\)
\end{eqnarray}
for the $\pm 1$  eigenvalue.
\par
Equations (4)-(5) represent a system of linear equations that  determine the elements $\chi_{ij}$ of the process $\mathcal{E}$.
Direct characterization can either be used to construct the complete process matrix or it may be applied partially to characterize only specific elements of $\chi$.
Partial characterization is especially useful when a priori knowledge about the quantum dynamics is available. 
For example, partial dynamics can determine the relaxation times $T_1,T_2$ efficiently by only characterizing $\chi_{\mathds{1}Z}$ and $\chi_{XY}$ \cite{Mohseni_PRL}.
\subsection{$[[6,0,2]]$ Concatenated Ancilla Code and Error Filtering}
\begin{table}[t!]
\centering
\begin{ruledtabular}
\begin{tabular}{cccccc}
      $i$ & $E_i$ & $e_i$ &$i$ & $E_i$ & $e_i$ \\[3pt]
      \hline
$0  $ & $ \mathds{1}\mathds{1}$ & $ 0  0  0  0  0  0$ & $8 $ & $ XY$ & $ 0   0   0   1   1   1 $ \\
$1 $ & $ X \mathds{1}$ & $ 0   0   0   1   0   0$ & $9 $ & $ XZ$ & $ 0   0   0   1   1   0 $ \\
$2 $ & $ Y\mathds{1}$ & $ 0   0   1   1   0   0 $ &$10 $ & $ YX$ & $ 0   0   1   1   0   1 $\\
$3 $ & $ Z\mathds{1}$ & $ 0   0   1   0   0   0 $ &$11 $ & $ YY$ & $ 0   0   1   1   1   1 $\\
$4 $ & $ \mathds{1} X$ & $ 0   0   0   0   0   1 $ &$12 $ & $ YZ$ & $ 0   0   1   1   1   0 $\\
$5 $ & $ \mathds{1} Y$ & $ 0   0   0   0   1   1 $ &$13 $ & $ ZX$ & $ 0   0   1   0   0   1 $\\
$6 $ & $ \mathds{1} Z$ & $ 0   0   0   0   1   0 $ &1$4 $ & $ ZY$ & $ 0   0   1   0   1   1 $\\
$7 $ & $ XX$ & $ 0   0   0   1   0   1 $ & $15 $ & $ ZZ$ & $ 0   0   1   0   1   0 $\\
\end{tabular}
\end{ruledtabular}
\caption{\label{tab:located_errors}
Error syndromes $e_i$ for states $\ket{i} = E_i \ket{0}$ indexed by the integer $i$ for the group of located errors $E_i \in \mathds{E}_{\bf P}$.
As evident through the one-to-one correspondence between the located error operators ($E_i$) and the syndromes ($e_i$), the code $\mathcal{S}_1$ is non-degenerate with respect to located errors
Errors involving weight-one ancilla operators ($\mathds{E}_{\bf P} \otimes \mathds{E}_{\bf A}  \in \mathds{E}$) are associated with error syndromes whose first two values are either $01,10$ or $11$. These errors are filtered out and do not affect the constructed $\chi_{mn}$.}
\end{table}
In its current form, DCQD assumes the ancilla system is noiseless and therefore any non-trivial syndrome measurements are attributed to the process acting solely on the principal system.
This provides a justification for interpreting the syndrome statistics in terms of the quantum process defined by Eq.~(\ref{eq:PM}) acting on the principal system.
Schemes assuming noiseless {\bf A} require the operator set $\mathds{E}$ have support only on {\bf P} in order to decode the syndrome. \cite{Mohseni_PRL,Omkar1}
Relaxing the assumption of ideal ancilla would introduce ambiguity into syndrome interpretation.
For example, in the Bell state previously used for DCQD \cite{Mohseni_PRL}, a stabilizer measurement cannot discriminate between a process that invokes no error on {\bf A} and a nontrivial error on {\bf P} and one that induces an error on {\bf A} while {\bf P} is unaffected.
\par
In practice, this ambiguity leads to errors in the characterization of the principal system as realistic ancilla also undergo quantum process.
Note that the error ambiguity, seen for example in Bell state DCDQ \cite{Mohseni_PRL}, comes from the invariance of the error syndrome under the interchange ${\bf A } \Leftrightarrow {\bf P}$.
To resolve this ambiguity, and differentiate between the different physical scenarios that lead to the same syndrome, we concatenate the ancilla {\bf A} qubits. 
Concatenation invalidates the mapping ${\bf A } \Leftrightarrow {\bf P}$, thus removing the syndrome ambiguity.
Concatenated ancilla qubits involve additional stabilizer generators such that the code detects low weight processes exclusive to system {\bf A}.
These new syndromes can be used  to filter the characterization data by rejecting those values that indicate errors on the ancilla.
This offers an improvement to a fundamental limitation of code-based process tomography. 
Moreover, the concatenated ancilla further partition the set $\mathds{E}$ into a set of {\em located} errors with support on only the principal system and a set of {\em unlocated} errors whose support is the composite system. 
\par
We now outline our main result, the construction of a code characterizing a $n_{\bf P}=2$ qubit principal subsystem with noisy ancilla.
The characterization code must first satisfy the located quantum Hamming bound $\sum_{j=0}^2 \binom{n_{\bf P}}{j} 2^k \leq 2^n$ \cite{Haselgrove}. 
The $k=0$ located Hamming bound (we choose $k=0$ since we wish to minimize overhead and are note interested in encoding logical information) is saturated for $n=4$, so we use the $[[4,0,2]]$ code $\mathcal{S}_0 = \<XIXI,IXIX,ZIZI,IZIZ\>$ to characterize {\bf P}.
However, as discussed in the last section, this code leads to the mistaken interpretation that processes acting on {\bf A} (qubits 3,4) characterize {\bf P} (qubits 1,2) under the interchange ${\bf A } \Leftrightarrow {\bf P}$ as is obvious from the symmetry of the generators.
\par
We can remove the syndrome degeneracy by encoding each physical qubit in {\bf A} with a second error detection code. 
This form of code concatenation enables the detection of processes that occur on only the ancilla.
Concatenation of the ancilla is also compatible with DCQD, as the first level stabilizers $\mathcal{S}_0$ are still used for direct process characterization. 
A schematic of this process is shown in Fig.~\ref{fig:schematic}.
The additional resources required for encoding the ancilla can be managed by adjusting the error detection properties of the second code. 
We encode the two ancilla qubits from the original characterization code with a $[[4,2,2]]$ code that is capable of detecting weight-one operators \cite{Gottesman_97,Terhal_RMP}. 
This brings the total number of qubits to six and forms a $[[6, 0, 2]]$ code.
\par
The encoding $[[4,2,2]]$ stabilizer group is $\mathcal{S}_E = \left \< XXXX,ZZZZ \right\>$, where we choose $\bar{X}_1 = X X II , \bar{Z}_1=ZI Z I, \bar{X}_2 = I X I  X, \bar{Z}_2 = I I ZZ$ as representative logical operators for the two encoded qubits.
Ancilla concatenation means replacing the ancilla qubits in $\mathcal{S}_0$ with the logical qubits from $\mathcal{S}_E$, that is, $X(Z)_{3,4} \mapsto \bar{X}(\bar{Z})_{1,2}$ and $\mathcal{S}_0 \mapsto \<XI\bar{X}\bar{I},IX\bar{I}\bar{X},ZI\bar{Z}\bar{I},IZ\bar{I}\bar{Z}\>$.
Expressed in terms of its generators, the newly formed code is 
\begin{eqnarray}
\label{eq:S_1}
\mathcal{S}_1 = &\<& IIXXXX, IIZZZZ, XIXXII,\\ \nonumber 
&& ZIZIZI, IXIXIX, IZIIZZ\>.
\end{eqnarray}
It is clear from Eq.~(\ref{eq:S_1}) that either one or both of the first two generators anti-commute with all weight-one errors on the four qubit ancilla subsystem. 
The remaining generators associate a unique error syndrome to {\em all} located errors on {\bf P} when {\bf A} is noisless.
Additionally, as detailed below, this code detects all errors occurring on the first two qubits {\bf P} simultaneous to any weight-one errors on {\bf A}.
\par
The group $\mathcal{S}_1$ stabilizes the (unnormalized) one dimensional codespace 
\begin{eqnarray}
\label{eq:codeket}
\ket{0} & = & \ket{000000}+\ket{001111}+\ket{010101} + \ket{011010} \\ \nonumber
&+ & \ket{100011} +\ket{101100} +\ket{110110}+\ket{111001}.
\end{eqnarray}
In addition, $\mathcal{S}_1$ partitions the Hilbert space into $64$ one-dimensional orthonormal subspaces $\mathscr{H}_i$, each of which is identified by a unique error syndrome $e_i$. 
\par
We now detail the set of errors $\mathds{E}$ for which QEC condition in Eq.~(\ref{eq:QECC}) is satisfied. 
We begin by structuring $\mathds{E}$ into two disjoint sets based on location of the induced errors or process.
The first set consists of located errors acting on the principal system {\bf P}, i.e., operators of the form $\sigma^i \sigma^j \mathds{1} \mathds{1} \mathds{1} \mathds{1}$.
We denote this set of 16 errors, which forms the Pauli two-qubit group modulus phases, by $\mathds{E}_{\bf P} \equiv \mathcal{P}_2/ \{\pm i, \pm1\}$.
It shall also be useful to refer to the 12 possible weight-one ancilla errors as $\mathds{E}_{\bf A}  \equiv  \{X_i,Y_i,Z_i\}$ where $i=3,4,5,6$ are the sites comprising {\bf A}.
The second set of errors, with 192 elements, consists of the tensor product of located and ancilla errors $\mathds{E}_{\bf P} \otimes \mathds{E}_{\bf A} $. 
Using the above definitions, the set of detectable processes is the disjoint union $\mathds{E} = \mathds{E}_{\bf P}\otimes \mathds{1}_{\bf A} \cup \mathds{E}_{\bf P} \otimes \mathds{E}_{\bf A} $.
\par
The noisy ancilla filtering properties of this code are evident upon inspecting the syndromes pertaining to the sets $\mathds{E}_{\bf P} \otimes \mathds{1}_{\bf A}$ and $\mathds{E}_{\bf P} \otimes \mathds{E}_{\bf A}$. 
The code is {\em non-degenerate} for the set of located processes, i.e., choosing $E_a,E_b \in \mathds{E}_{\bf P} $ the code matrix in Eq.~(\ref{eq:QECC}) becomes $C_{ab}=\delta_{ab}$ for the state in Eq.~(\ref{eq:codeket}).
Elements $E_i \in \mathds{E}_{\bf P} $ map the codestate to distinct orthogonal states $\ket{i} = E_i \ket{0}$, with distinct syndromes $e_{i}$ for $i \in \[0,15\]$.   
This group of located processes commutes with the first two generators in Eq.~(\ref{eq:S_1}), so the corresponding syndromes are of the form $e_i=(0,0,e_{i3},e_{i4},e_{i5},e_{i6})$.
Table \ref{tab:located_errors} enumerates the syndromes associated with all located errors.
Syndromes that begin with ``00" indicate {\bf A} is error free and that the corresponding measurement is accurate for characterizing $\chi$ as described in Eqs.~\ref{eq:syn_prob_diag},\ref{eq:syn_prob_coherence},\ref{eq:syn_prob_coherence_2}.
\par
The code is degenerate for processes $E_a, E_b \in \mathds{E}_{\bf P} \otimes \mathds{E}_{\bf A} $.
This result is expected since there are $2^6=64$ syndromes and $208$ operator elements in the set $\mathds{E}$.
The remaining 192 processes $E_j \in \mathds{E}_{\bf P} \otimes \mathds{E}_{\bf A} $ ($j \in \[16,207\]$) map the codeword onto the remaining 48 orthogonal states, $\ket{i}$ with $i \in \[16,63\]$.
Those processes that have an odd weight support on {\bf A} anti-commute with either one or both of the first two generators of $\mathcal{S}_1$.
Consequently, syndromes that begin with $01,10,11$ indicate that noise was detected on the ancilla. 
Because these syndromes are degenerate we cannot know exactly which process corrupted the ancilla.
Therefore, this data is {\em filtered} out from characterizing the principal system.

\section{Numerical Characterization of Amplitude Damping Channel} 
\label{sec:Monte-Carlo}
\begin{figure}[htbp]
\begin{center}
\includegraphics[width=\columnwidth]{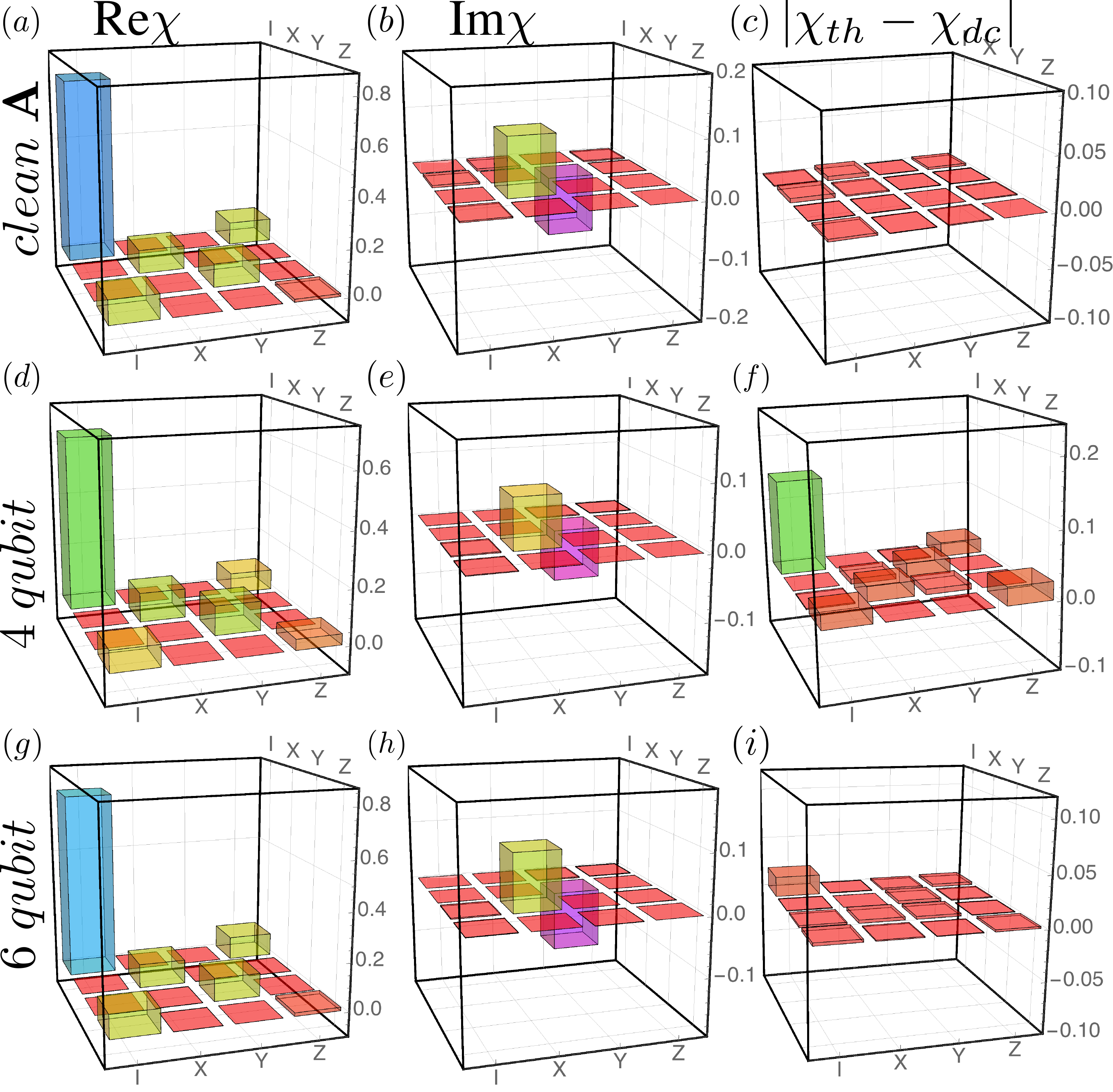}
\caption{\label{fig:chi}
Simulated AD channel process matrices constructed from ensembles of syndrome measurements according to the DCQD procedure, namely Eqs.~\ref{eq:syn_prob_diag},\ref{eq:syn_prob_coherence},\ref{eq:syn_prob_coherence_2}. 
Probabilities are determined from $10^6$ Monte-Carlo events using the numerical parameters $\gamma=.4$, $p=.1$ were used for the amplitude damping and depolarizing channels respectively.  
The real and imaginary parts for $\chi$ constructed with a noiseless ancilla is given in panels (a,b) and its difference from the theoretical value appears in panel (c).
A noisy ancilla reduces the accuracy of the DCQD procedure as seen in panels (d-f) for which a standard $[[4,0,2]]$ has been used. 
As seen in panels (g-i), weight-one ancilla errors are mitigated by utilizing a concatenated $[[6,0,2]]$ code. 
The constructed $\chi$ matrix is characterized by a high degree of fidelity, $F(\mathcal{E}^{AD}_{\bf P} (\rho),\mathcal{E}^{[[6,0,2]]}(\rho))=.9884$.  } 
\end{center}
\end{figure}

We now test the procedure outlined in the previous section (and Fig.~\ref{fig:DCQD_flowchart}) by numerically constructing the process matrix for the well known amplitude damping (AD) channel.
While we could characterize arbitrary noise on {\bf P}, for clarity we consider the case when only the first qubit experiences AD. 
The principal system process matrix is $\mathcal{E}^{AD}_{\bf P} (\rho) =  \sum_{a} E_a^{AD} \rho E_a^{AD\dagger}$ where AD channel is written in terms of the Kraus operators $E_0^{AD} = (1 + \sqrt{1-\gamma})\mathds{1}/2 +  (1 - \sqrt{1-\gamma}) Z_1/2, E_1^{AD} = \sqrt{\gamma} (X_1+i Y_1)/2$. 
Expressing the $\chi$ matrix in the Pauli basis $F_i=\{I, X, Y, Z\}$ the only non zero process matrix elements appear along the diagonals $\chi_{II}= (1 + \sqrt{1-\gamma})^2/4, \chi_{XX} = \chi_{YY} = \gamma /4,\chi_{ZZ}= (1 - \sqrt{1-\gamma})^2/4$ and anti-diagonals $\chi_{IZ} = \chi_{ZI} = \gamma/4, \chi_{YX} = - \chi_{XY} = i\gamma /4$ (see Fig.~\ref{fig:chi}).
To test our code in the presence of a noisy {\bf A} subsystem we take the state $\mathcal{E}^{AD}_{\bf P} (\rho) $ and subject it to an additional depolarizing (DP) channel acting independently on each ancilla qubit.
We construct the channel via a composition of single qubit DP channels so that $\mathcal{E}_{\bf A}^{DP}(\rho) = \mathcal{E}_3^{DP}(\rho) \circ \mathcal{E}_4^{DP}(\rho) \circ \mathcal{E}_5^{DP}(\rho) \circ \mathcal{E}_6^{DP}(\rho)$ where $\mathcal{E}_i^{DP} (\rho) = (1-p) \rho + p (X_i \rho X_i + Y_i \rho Y_i + Z_i \rho Z_i)/3$ and  $ f(\rho) \circ g(\rho) = f ( g (\rho))$ denotes the usual functional composition of mappings.
The resulting state is $\mathcal{E}^{DP}_{\bf A}\(\mathcal{E}^{AD}_{\bf P} (\rho) \)$ where for order of the the independent {\bf P}, {\bf A} channels is arbitrary. 

To simulate experimental measurement statistics we perform a Monte-Carlo simulation in which we project the state $\mathcal{E}^{DP}_{\bf A}\(\mathcal{E}^{AD}_{\bf P} (\rho) \)$ into the $\pm 1$ eigenstate of each generator in Eq.~\ref{eq:S_1} with probability $\text{Tr} \[ (1\pm g_l) \mathcal{E}^{DP}_{\bf A}\(\mathcal{E}^{AD}_{\bf P} (\rho) \)\]$. 
This procedure is repeated for all six syndromes with the $\pm 1$ eigenvalues for each generator defining a single measured syndrome generator. 
The probabilities for each clean syndrome, i.e., ``00" syndromes, with respect to all clean results is used to determine the $\chi$ elements by Eqs.~(\ref{eq:syn_prob_diag}), (\ref{eq:syn_prob_coherence}), and (\ref{eq:syn_prob_coherence_2}).
Following this procedure we perform a Monte-Carlo simulation of the following three scenarios: 
(i) the AD channel $\mathcal{E}^{AD}_{\bf P} (\rho)$ acting on qubit 1 with a noiseless ancilla {\bf A} ($\mathcal{E}_{\bf A}^{DP}(\rho)= \rho$), 
(ii) AD on qubit 1 with a noisy {\bf A} implemented with detection being done by a non-concatenated $[[4,0,2]]$ DCQD code and 
(iii) AD on qubit 1 with a noisy {\bf A} using the $[[6,0,2]]$ code given in Eq.~\ref{eq:S_1} to determine $\chi$.

The process matrix constructed in scenario (i), i.e. for a noiseless ancilla system, is shown in Fig.~\ref{fig:chi} panels (a,b) and is compared to the theoretical result in panel (c). 
Finite sampling causes a small discrepancy between the theoretical and the simulated result as seen in panel (c). 
Next, we simulate case (ii) involving the four qubit non-concatenated DCQD code in which every possible syndrome is used to determine the $\chi$. 
The absence of a filtering process means that each error occurring on the ancilla system corrupts the syndrome probabilities which, in turn, determine $\chi_{i,j}$. 
The simulated $\chi$ matrix is presented in Fig.~\ref{fig:chi} panels (d,e) and its distance from the clean $\chi$ is given in panel (f).
Finally, for case (iii), we construct $\chi$ using the ancilla concatenated code (Eq.\ref{eq:S_1}) and present the results in panels (g-i).
Inspecting panels (c), (f), and (i) it is obvious that the $\chi$ matrix constructed with the concatenated code is more accurate than the standard non-concatenated code. 

To quantify this difference we calculate the fidelity,defined as $F(\rho,\sigma) = \text{Tr}[\sqrt{\sqrt{\rho}\sigma\sqrt{\rho}}]$ for two density matrices $\rho, \sigma$, between the theoretical and numerically constructed $\chi$ coefficients. 
The states we calculate the fidelity of are one qubit states subjected to our constructed amplitude damped channels and the theoretical channel, that is: $\mathcal{E}^{[[6,0,2]]}(\rho)= \sum_{mn} \chi_{mn}^{[[6,0,2]]} F_m \rho F_n^\dagger$, $\mathcal{E}^{[[4,0,2]]}(\rho)= \sum_{mn} \chi_{mn}^{[[4,0,2]]} F_m \rho F_n^\dagger$, and  $\mathcal{E}^{AD}_{\bf P} (\rho)$.
Using an initial single qubit state $\rho = \ket{0}\bra{0}$ we find $F(\mathcal{E}^{AD}_{\bf P} (\rho),\mathcal{E}^{[[6,0,2]]}(\rho))=.9884$ and $F(\mathcal{E}^{AD}_{\bf P} (\rho),\mathcal{E}^{[[4,0,2]]}(\rho))=.9165$ which represents a 10\% improvement in the fidelity for the constructed process matrix for the specific case of $p=0.1$.
\begin{figure}[tbp]
\begin{center}
\includegraphics[width=8cm]{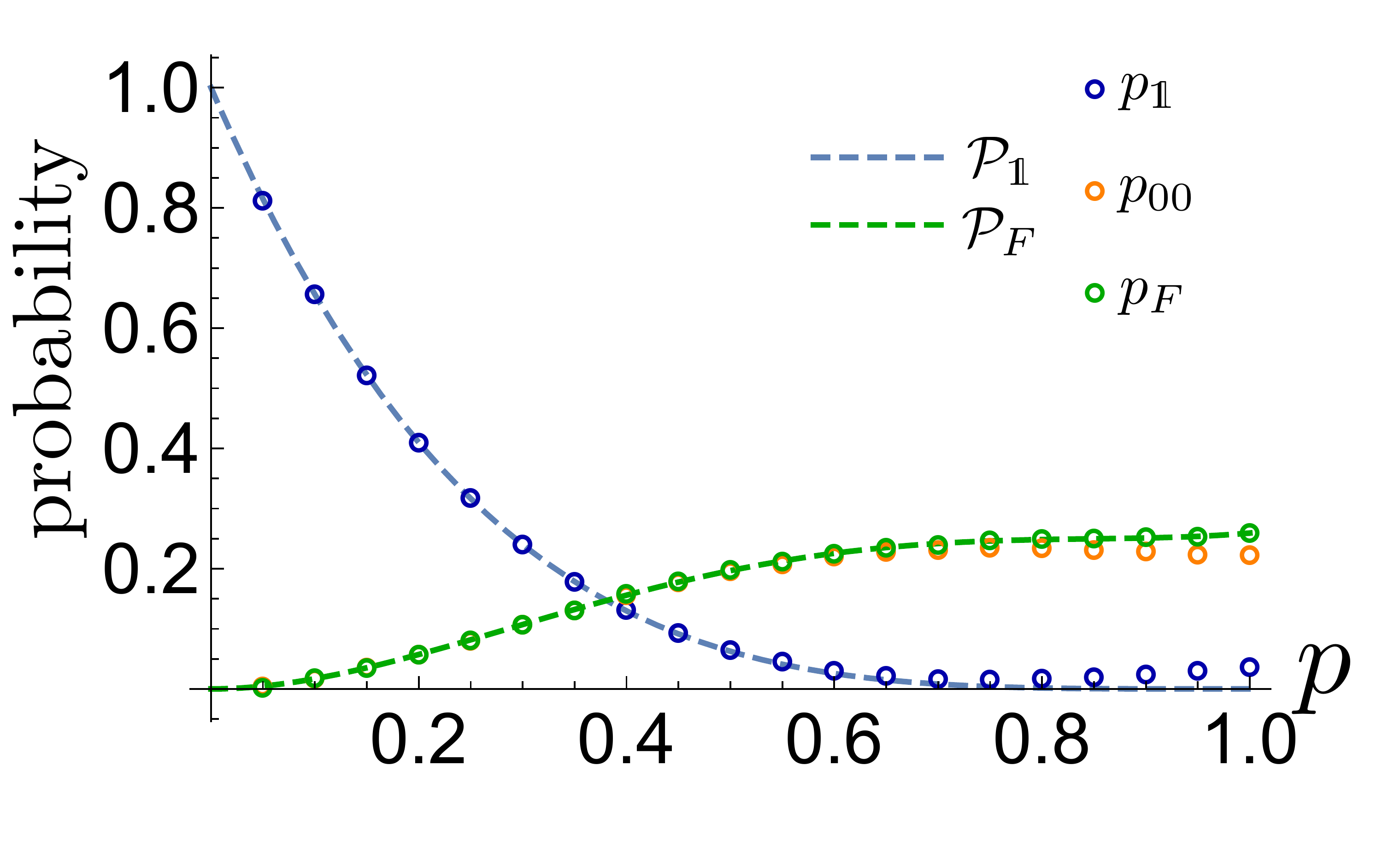}
\caption{\label{fig:failure}
Probabilities of located error syndrome to occur in the presence of a noiseless principal system {\bf P} in the presence of independent depolarizing channels acting on the ancilla {\bf A} with probability $p$. 
The blue dashed line ($\mathcal{P}_\mathds{1} = (1-p)^4$) shows the probability for the identity operator to occur while the probability for the identity syndrome to appear in the Monte-Carlo simulation is given by the blue circles (denoted by $p_\mathds{1}$). 
Stabilizer operators therefore occur with probability $\Delta p_{\mathds{1}} = p_\mathds{1}- \mathcal{P}_\mathds{1}$.
Orange circles ($p_{00}$) denote the rate at which the remaining located (``00") syndromes occur.
Green circles (dashed lines) denote the simulated (theoretical) failure rate $p_F = p_{00} + \Delta p_{\mathds{1}}$ ($\mathcal{P}_\mathds{1} = p_2/3 + 2p_3/9+ 21p_4/81$, see text for derivation) indicating a located error syndrome from Tab.~\ref{tab:located_errors} is caused by an operator different than $E_i$. }
\end{center}
\end{figure}
\section{Filtering Failure Rate}
\label{sec:analysis}
Figure~\ref{fig:chi} shows how encoded ancilla can improve process characterization as compared to previous QEC-based schemes. 
This works by detecting noisy ancilla operations and filtering out those measurements, thus improving the accuracy of the constructed process matrix. 
We can explain the improvements in characterization fidelity in terms of a gain in the signal to noise ration for the characterization process.
In particular, we define the signal as the constructed process matrix elements, which are directly related to a measured syndrome $e_i$.
Noise is then those syndromes in which an error $E_j$ occurs on {\bf P} but we measure a syndrome $e_i$ with $j\neq i$.
The rate at which this occurs is represented by $p_F$ in Fig.~\ref{fig:failure}. 
To quantify the improvement in the signal to noise ratio we now analyze a simple model, the DP channel $\mathcal{E}_{\bf A}^{DP}(\rho)$, and compare the probability of failure for concatenated and non-concatenated characterization codes.
\par
Using the definition of noise, we say that the filter fails if data collected from a noisy event is used to characterize $\chi_{mn}$. 
In the $[[6,0,2]]$ code, failure cannot be due to any weight-one errors on {\bf A} since they, and their product with all located errors ($\mathds{E}_{\bf P} \otimes \mathds{E}_{\bf A}$), are within the detectable errors set $\mathds{E}$. 
The filter does however fail in the presence of some weight-two errors which commute with the first two generators in Eq.~\ref{eq:S_1}.
In the DP channel $\mathcal{E}_{\bf A}^{DP}(\rho)$ the probability for a weight $j$ to occur is $p_j=(1-p)^{4-j}\binom{4}{j}p^j$. 
Notably, the probability for the weight 0 ``error" $\mathds{1}$ to occur is the probability that the identity occur on each qubit $p_{0} = (1-p)^4$ which appears as the dashed blue line labeled $\mathcal{P}_{\mathds{1}}$ in Fig.~\ref{fig:failure}. 
Ancilla errors outside the correctable error set $\mathds{E}_{\bf A}$ occur with probability $p_{\geq 2} = p_2+p_3+p_4$.
However, not all of the errors with weight $\geq 2$ will lead to faulty characterization data since many most of them will still lead to syndromes beginning with one of $01,10,11$ and therefore do not corrupt the constructed $\chi$.
In these cases we discard the data point because it is (correctly) assumed that some error has occurred on {\bf A}. 

To confirm our estimates, we numerically calculate the failure rate with $10^6$ Monte-Carlo simulations of the composite depolarizing channel $\mathcal{E}_{\bf A}^{DP}(\rho)$.
With a single exception, the failure probability is by definition the number of syndromes beginning with $00$ divided by the total number of randomly generated errors. 
The exception comes from the ambiguity of whether the syndrome $e_0=\{0,0,0,0,0,0\}$ should count towards the error rate, as $e_0$ may be generated by the identity mapping or by any element in the normalizer group $\mathcal{N}(\mathcal{S}_1)$, i.e. the group of errors commuting the all stabilizer elements.

However, we know that the identity operator ($\mathds{1}^{\otimes 4}$) occurs with probability $\mathcal{P}_{\mathds{1}} = (1-p)^4$ as illustrated by the blue dashed curve in Fig.~\ref{fig:failure}. 
We determine the rate for erroneous identity-like syndromes to be $\Delta p_{\mathds{1}} = p_\mathds{1}- \mathcal{P}_\mathds{1}$, the difference between $\mathcal{P}_{\mathds{1}}$ and the numerical rate at which we measure the identity syndrome (blue circles in Fig.~\ref{fig:failure}). 
In Fig.~\ref{fig:failure} the green circles represent the total failure rate obtained by adding the identity probabilities difference to the probability with which all other located syndromes occur. 
Enumerating the number of weight 2,3, and 4 errors which commute with the first two generators of $\mathcal{S}_1$ and the probability with which they occur we find the probability of failure to be $\mathcal{P}_F=2 p_2/3 + 2p_3/9+ 21p_4/81$ where $p_{2,3,4}$ is the for probability for an error of weight 2,3, or 4 to occur. 
This function of $\mathcal{P}_F$ is plotted as the dashed green line in Fig.~\ref{fig:failure} and exactly matches our numerical data.  
The leading term in $\mathcal{P}_F$ goes as $O(p^2)$ in contrast to to non-concatenated DCQD schemes whose failure rate gores as $O(p)$, the probability for weight-one errors. 
 explains the sharp contrast in the constructed process matrices in the second and third rows of Fig.~\ref{fig:chi}.

\section{Discussion}
\label{sec:conclusion}
We have introduced a DCQD code that directly characterizes the quantum dynamics of a principal system with assistance from a noisy ancilla system.  
Within the stabilizer framework, we show that ancilla noise can be distinguished from processes acting on the principal system by using syndrome value as a filter for non-trivial ancilla processes.
For the example of DCQD with a $[[4,2,2,]]$, we have concatenated the ancilla qubits for purposes of detecting weight-one processes.
and compared the characterization of an amplitude damping process on the principal system using three different approaches: (i) clean ancilla system, (ii) noisy ancilla using a standard DCQD, and (iii) noisy ancilla using our concatenated $[[6,0,2]]$ code. 
Our numerical simulations found that the process matrix constructed using the six-qubit code shows a marked improvement in fidelity over the non-concatenated approaches.
\par
Our motivation for encoding the ancilla qubits has been to filter out those measurements that correspond to unwanted data. 
From this perspective, ancilla encoding represents a form of filtering the dynamics to isolate non-trivial processes acting only on the principal system.
We have argued that filtering increases the signal-to-noise ratio for process characterization, as measured by the gain in fidelity of the constructed matrix.
Of course, the gain for process characterization depends strongly on the details of the ancilla filter.
For example, the 6-qubit code introduced here detects only weight-one ancilla errors and their product with located principal system errors $\mathds{E}_{\bf P} \otimes \mathds{E}_{\bf A}$. 
When higher weight errors are common, the benefit of this ancilla encoding diminishes, and larger distance codes are needed to filter higher weights processes.
For example, a distance 4 code that detects all weight-2 ancilla errors will have a filter failure rate that scale as $O(p^3)$ with $p$ the ancilla error rate.
We could also have used a non-degenerate $[[5,1,3]]$ code to encode the ancilla, where each detectable error would have a unique error syndrome. 
In this case, each syndrome would be used without a filtering procedure.
In general, one can improve the signal to noise ratio at the expense of additional ancilla qubits and larger codes. 
\par
Additionally, we have taken $k=0$ throughout thought this work, but we could have used a $k\neq0$ code satisfying a generalized Hamming bound \cite{Haselgrove}. 
For example, a non-concatenated code performing error correction on two qubits with another encoded is provided in Ref.~\citenum{Omkar1}.  
Equations (\ref{eq:syn_prob_diag})-(\ref{eq:syn_prob_coherence_2}) are easily generalized using the higher dimensional projectors $Pi_i$
resulting in a code which detects ancilla errors while encoding some non-trivial quantum information. 
\par
Recent progress in realizing stabilizer QEC circuits with 9 and 4 qubits on different lattice configurations suggest that the implementation of these ideas should be experimentally feasible in the near future \cite{Martinis_15,IBM_15}. In particular, it is worth noting that the characterization processes described here and in earlier DCQD work do not require active, feed-forward error correction for purposes of implementation. Consequently, the use of QEC-based DCQD appears to be a natural way point toward the demonstration of error-corrected computation.
\section*{Acknowledgments}
E. D. and T. S. H. acknowledge support from the Intelligence Community Postdoctoral Research Fellowship Program. 
This manuscript has been authored by UT-Battelle, LLC, under Contract No. DE-AC0500OR22725 with the U.S. Department of Energy. 
The United States Government retains and the publisher, by accepting the article for publication, acknowledges that the United States Government retains a non-exclusive, paid-up, irrevocable, world-wide license to publish or reproduce the published form of this manuscript, or allow others to do so, for the United States Government purposes. 
The Department of Energy will provide public access to these results of federally sponsored research in accordance with the DOE Public Access Plan.

\end{document}